\documentclass[letterpaper,twocolumn,10pt]{article}
\pdfoutput=1
\usepackage{usenix,epsfig,endnotes}
\usepackage{enumerate}
\usepackage{color}
\usepackage{graphicx}
\usepackage{subcaption}
\usepackage{amsmath}
\usepackage{amssymb}
\usepackage{booktabs}
\usepackage{enumitem}
\usepackage{tabularx}
\usepackage{import}
\usepackage{hyperref}
\usepackage{multirow}
\usepackage{makecell}
\usepackage{appendix}
\usepackage{adjustbox}
\usepackage{array}
\hypersetup{
  pdfinfo={
    Title={FIRED: Frequent Inertial Resets with Diversification for Emerging
      Commodity Cyber-Physical Systems},
    Author={Miguel A. Arroyo, Hidenori Kobayashi, Simha Sethumadhavan, Junfeng Yang}
  }
}
\begin{document}

\date{}

\title{\Large \bf FIRED: Frequent Inertial Resets with Diversification\\for Emerging Commodity Cyber-Physical Systems}
\author{
  {\rm Miguel A. Arroyo}\\
  \and
  {\rm Hidenori Kobayashi}\\
  \and
  {\rm Simha Sethumadhavan}\\
  \and
  {\rm Junfeng Yang}\\
  \and
  Columbia University\\
  \{miguel,hidenori,simha,junfeng\}@cs.columbia.edu
}

\maketitle

\def\pname{FIRED}
\def\R{Reset}
\def\r{reset}

\subsection*{Abstract} A Cyber-Physical System (CPS) is defined by
its unique characteristics involving both the cyber and physical
domains. Their hybrid nature introduces new attack vectors, but
also provides an opportunity to design new security defenses. In
this paper, we present a new domain-specific security mechanism,
\pname{}, that leverages physical properties such as inertia of the
CPS to improve security.

\pname{} is simple to describe and implement. It goes through two
operations: \R{} and Diversify, as frequently as possible -- typically
in the order of seconds or milliseconds. The combined effect of
these operations is that attackers are unable to gain persistent
control of the system. The CPS stays safe and stable even under
frequent \r{}s because of the inertia present. Further, \r{}s
simplify certain diversification mechanisms and makes them feasible
to implement in CPSs with limited computing resources.

We evaluate our idea on two real-world systems: an engine management
unit of a car and a flight controller of a quadcopter. Roughly
speaking, these two systems provide typical and extreme operational
requirements for evaluating \pname{} in terms of stability, algorithmic
complexity, and safety requirements. We show that \pname{} provides
robust security guarantees against hijacking attacks and persistent
CPS threats. We find that our defense is suitable for emerging CPS
such as commodity unmanned vehicles that are currently unregulated
and cost sensitive.

\section{Introduction}

A Cyber-Physical System (CPS) represents the synthesis of computational and
physical processes encompassing a wide range of applications including
transportation, medical, robots, and power grids. CPSs usually consist of a
controller, sensors, actuators and a network. Known attacks, academic and
real-world, on CPSs are diverse and span both the cyber and physical domains.
Cyber attacks like the Jeep Hack by Miller et al. \cite{miller2015remote} and
Tesla remote hack by the Keen group \cite{keen2016tesla} and physical attacks
such as those on car sensors by Yan et al. \cite{yan2016vehicles} are just a few
that have appeared in the media recently. The dangers were first highlighted by
the pioneering work of Koscher et al. \cite{koscher2010experimental} and
Checkoway et al. \cite{checkoway2011comprehensive}.

Current state of the art defenses for CPSs can be understood as defenses for the
physical portion of the CPS, viz., the sensors, and cyber portion of the CPS,
viz. the control and communication software. The physical defenses leverage
physical properties to detect and mitigate sensor spoofing, sensor jamming, and
other similar types of attacks. Current cyber defenses, typically designed for
more traditional systems, are usually incompatible for a CPS due to timing or
computing constraints but have been adapted by modifying some of the operational
parameters.

In contrast to current cyber defenses for CPSs, which are adaptations
of cyber defenses, in this paper, we propose a new software security
defense for CPS that leverages unique properties of CPSs. A key
innovation in our approach is that we take advantage of inertia,
i.e., the ability of the CPS to stay in motion or at rest, and its
ability to tolerate transient imperfections in the physical world,
to survive attacks. Basically we reset the controller software as
frequently as possible without impacting safety. The security benefit
of the frequent reset defense is that it limits the time for which
the system is vulnerable to attacks. Additionally, diversification
is used to force the adversary to develop a new attack strategy
after every \r{}; that is, attacks after each \r{} should be
independent of each other, lowering the likelihood of an attack's
success.

Our system, \pname{}, is best understood with an example. Consider
a drone: even if power is cut off to the engine, the drone will
continue to glide due to inertia; similarly, even if a few
sensor inputs are incorrect, the drone will continue to operate
safely because intermittent sensor errors happen in normal operation
and controllers are designed to handle this case. In \pname{}, we
take advantage of these features. We intentionally \r{} the system
periodically to clear state that may have been corrupted by an
attacker. During \r{}s, we rely on system inertia to continue
operating, and we diversify to prevent the same attacks from breaking
the system.

It is easy to see why \pname{} is uniquely suited to CPSs as a
defensive technique. If traditional digital, non-physical systems
are reset frequently, because they lack inertia, they cannot
independently survive resets. To wit, consider \r{}ing a server
every few network packets. In order to recover from missed packets,
an external entity needs to retransmit them. In contrast, in a CPS,
the feedback loops present help them independently rebuild the state
that might have been lost due to the reset. Additionally, the
physical components of the system, due to inertia, continue operating
even when there is no computation by the controller. Finally, the
network packets can arrive in the order of nanoseconds in a traditional
digital, non-physical system while the \r{} times are in the order
of seconds. For CPS data often arrives on human scales (milliseconds)
and reboot times are also much shorter.

Realizing \pname{} for CPSs is not without challenges. While all CPSs have
inertia, the amount of inertia they have is variable. \R{}ting too frequently
may cause the safety requirements of the system to be violated if the inertia
cannot cover the operational and safety requirements during controller downtime
during \r{}. Further, computational resources are constrained on most CPSs due
to cost and power requirements. In this paper we show a systematic method to
determine the frequency of \r{} and an implementation of \pname{} that allows
CPS to be secured in a pragmatic way.

We evaluate \pname{} on two popular CPSs with different inertia and safety
requirements. Using an open source engine controller---rusEFI---and measurements
from a real combustion engine, we discuss and measure the performance and safety
impacts. Additionally, we evaluate \pname{} on a Flight Controller (FC) for a
quadcopter which exhibits more limited inertia. We find that both of these
systems tolerate multiple frequent \r{}s. The engine can tolerate \r{}s at
intervals of 125ms while the quadcopter at an interval as frequent as a second.
We additionally define appropriate
safety requirements for each and show that \pname{} does not violate them. Thus,
\pname{} provides a novel way for enhancing the security of CPSs without
impacting safety, with low resource requirements, and a simple implementation
using unique properties of CPSs.

The rest of the paper is organized as follows: In the next section we
describe a system model for a CPS system that we use, in the following
section we describe the attack model. We then describe our \pname{}
technique in Section~\ref{sec:model}, provide security and safety
analysis in Section~\ref{sec:model-analysis}, and evaluate \pname{}
on representative case studies---an Engine Control Unit (ECU) and
a Flight Controller (FC)---discussed in Section~\ref{sec:evaluation}.
We conclude in Section~\ref{sec:conclusion} with applicability and
limitations.

\section{System Model \label{sec:system-model}}

\begin{figure} \centering
  \includegraphics[width=0.9\linewidth]{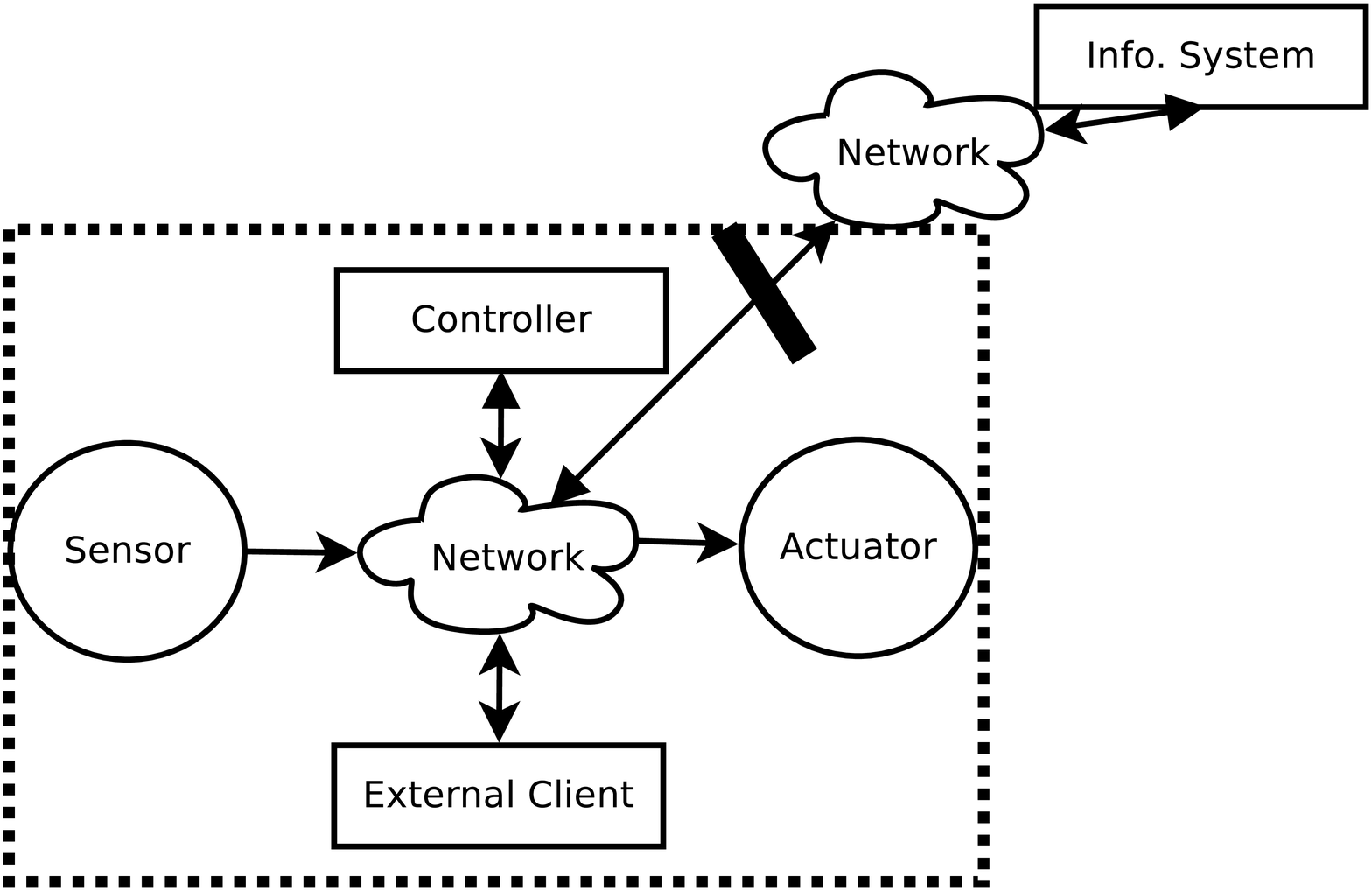}
  \caption{CPS System Model. The portion enclosed in the dotted area corresponds
    to the safety critical components. \label{fig:cps-model}}
\end{figure}

\noindent\textbf{Overall Architecture:} Current CPSs have four main
subsystems: a subsystem that interfaces with the physical world
that includes the sensors and the actuators, a subsystem that takes
inputs from the sensors and generates commands for the actuators---the
control subsystem,---a subsystem that takes in external commands,
and finally a subsystem the provides non-critical functions such
as entertainment subsystem. In most CPSs today these subsystems are
not isolated from each other creating a quagmire of security problems.
However, recognizing these dangers, emerging CPS designs include
primitives for isolation (such as separating out the networks in a
car), sensor authentication and encryption to name a few.

We set our work in this emerging model and we consider the safety
and security of only the software that provides control over the
physical system (i.e. the safety critical components). The enclosed
portion of Figure~\ref{fig:cps-model} illustrates our model. The
rationale for this choice is to focus our attention on aspects that
are unique to CPS. A wide variety of systems fall into this model,
such as unmanned transport vehicles or industrial control systems.

\textbf{Resource Model:} CPS resources can vary wildly. However, a large subset
of these systems rely on microcontrollers. They tend to execute from ROM or Flash
chips with limited memory (typically hundreds of KBs or a few MBs), have no
memory management unit (MMU), only a memory protection unit (MPU); and the
performance offered by these microcontroller-based systems is orders of
magnitude lower compared to traditional microprocessor-based systems such as
servers and smartphones. The primary driving factor for the system design
choices is the cost of components, and a need to provide real-time deterministic
execution characteristics which preempts use of speculative techniques for
achieving high performance. As a result of these stringent resources and
requirements, these systems usually run a library operating system---commonly
referred to as Real-Time OS (RTOS)---variant\endnote{In a library OS, the
application is compiled with a library that provides typical OS functions and
hardware abstractions.}, and porting of existing security defenses that weren't
designed for these low-resource devices can be difficult or sometimes
impossible. Similarly, these architectural or platform differences present
challenges to mounting certain classes of attacks. For instance, modifying even
a small amount of data, say one byte, requires an entire Flash sector to be
modified at once, that incurs latencies in the order of hundreds of milliseconds
to seconds. So persistent attacks that need to modify Flash can take several
seconds while the same can be accomplished in hundreds of nanoseconds on
traditional systems -- nearly seven orders of magnitude difference.

\noindent\textbf{CPS Properties:} With these system and resource models in mind,
we can begin to discuss some of the important properties of cyber-physical
systems that allow for interesting security techniques to be explored. These
properties stem from the design, algorithms, and physics of CPSs. The main
properties of CPSs are:

\vspace{0.04in}
\noindent\textit{\textbf{They move.}} CPSs operate in the physical
domain and as a result involve some form of lateral or rotational
motion. This motion leads these systems to have \emph{inertia}--the
resistance of an object to any change in its motion. This is essential
for \pname{} as it asserts that these systems can continue operation
and exhibit a tolerance to missed events, either by design or due
to faults.

\vspace{0.04in}
\noindent\textit{\textbf{Environment can be observed.}} The fact that
CPSs can observe their environment using sensors is fundamental in
their design and operation. The inclusion of sensors also means
that any actuation of the system can be observed by the sensors.
This further implies that the state of the system that caused the
actuation can be relearned by the sensors. Roughly speaking this
is the same as storing the state of the system (before actuation)
to a data store outside the system (the environment), and then
reading it back to the system (through the sensors). This feature
is critical for \pname{} as it allows for certain state to be
discarded from the system during resets, i.e., transmitted out of
the system before a reset, and re-observed once out of reset. The
quality of state may be degraded in this process but as we will see
next, CPSs are, by design, resilient to modest degradation.

\vspace{0.04in}
\noindent\textit{\textbf{Imperfections are tolerable.}} The physical
quantities that need to be sensed are continuous where as digital
systems are discrete. This mismatch requires conversion from continuous
data to discrete data which introduces quantization error -  as a
concrete example, while quantities such as pressure, temperature
etc. can in theory have infinite precision, the precision of the
readings is limited by the analog-to-digital converter found in
these sensors which is typically 8-12 bits. Thus the inputs to a
CPS are approximations of the real world. Furthermore, sensors
also have accuracy problems as the sensing mechanism can (and does)
provide a different response depending on environmental conditions,
not all of which can be completely characterized. Also, physical
artifacts in the environment (such as flying close to an electric
line) can perturb sensors. As a result, algorithms and hardware
of a CPS are typically designed to account for modest levels of
uncertainty. Thus, these systems include multiple sensors from which
physical world measurements can be estimated. Furthermore, the
algorithms are capable of self-correction. As we will see later
this property will allow \pname{} to operate smoothly.

\section{Adversary Model \label{sec:adversary-model}}

\noindent \textbf{Attack Surfaces:} The emerging CPS model considered in this
paper has three attack surfaces:

\vspace{0.04in}
(1) \textit{\textbf{Sensors \& Actuators:}} The CPS's interface to the
physical world through sensors and actuators. This is susceptible to physical
threats such as tampering of sensors and/or actuators, sensor spoofing and
jamming.

\vspace{0.04in}
(2) \textit{\textbf{Software:}} The control software that handles incoming
queries, processes sensor data, and computes actuator commands. This is
susceptible to the more traditional software threats such as those stemming from
memory vulnerabilities, integer overflows, etc.

\vspace{0.04in}
(3) \textit{\textbf{Network:}} The network that connects the various
components to the controller. This is susceptible to threats such as
man-in-the-middle type attacks.
\vspace{0.04in}

For this work, we consider an adversary that seeks to exploit the control
software for these CPSs. We point out that physical-subsystem attacks may
serve as entry points to exploit the control software.

\textbf{Physical-subsystem Attacks:} Physical attacks typically target state
estimation and control; in fact, much work has been done on this front. Attacks
on state estimation usually manifest themselves in one of two ways: sensor
spoofing and sensor jamming. The difference between these two threats is
effectively in the level of control that can be exerted over the sensor, with
spoofing being more sophisticated than jamming. Examples of these works are
presented by Kerns et al. \cite{kerns2014unmanned} where they spoof GPS signals
to capture and control unmanned aircraft, Davidson et al.
\cite{davidson16controlling} who describe how to spoof optical flow sensors in
quadcopters, Kune et al. \cite{kune2013ghost} that present how electromagnetic
interference can be used to attack sensors, and Son et al. \cite{son2015rocking}
which incapacitates a drone's gyroscopes by using intentional noise. It is
also important to note that while these attacks are physical in nature, they can
affect the cyber component of the system or provide the opportunity for certain
cyber attacks to become effective. Also, jamming and spoofing attacks require a degree of spatial proximity: the jamming or spoofing device needs to keep up
with the motion of the CPS for continued action.

\textbf{Cyber-subsystem Attacks:} Attacks in the cyber portion of a CPS are
essentially the same as those seen on more traditional systems. For example,
memory vulnerabilities, such as code reuse and data corruption, are as much a
problem for CPS as for other systems. Work by Checkoway et al.
\cite{checkoway2011comprehensive} and Koscher et al.
\cite{koscher2010experimental} provide a CPS specific discussion on some of
these threats. As discussed above, physical threats can indirectly trigger certain cyber
vulnerabilities, for example, integer overflows and underflows. Maliciously
manipulated sensor values can cause incorrect branches to be executed in the
control algorithms, or worse trigger CPU specific vulnerabilities such as those
discussed by Rosenberg \cite{rosenberg2014qsee} for the ARM TrustZone.

\textbf{Goals and Capabilities:} We make specific assumptions about
an adversary's capabilities.

\vspace{0.04in}
  \textit{\textbf{Attacker's intention is to gain a persistent foothold
into the system.}} Achieving persistence involves compromising at least one of
the controllers available. By gaining a persistent foothold, an adversary can hijack the
targeted system and attack at a time and place of their choosing. In
other words, an attacker's immediate objective is to sacrifice the integrity of
the system. Prior work has considered availability
attacks~\cite{sakr2015security} and the defenses provided here are orthogonal to
prior defenses.
\vspace{0.04in}

\textit{\textbf{An attacker will avoid causing irreversible harm to the
          system.}} Since an attacker's objective is to persist in the system
      any irreversible damage may compromise their goals. As a result, an
      attacker will avoid inducing fatal failure modes (i.e. destruction of the
      system).

      \vspace{0.04in}

      \noindent\textit{\textbf{An attacker has complete knowledge of the system
internals.}} The physics of the system and the control algorithms used are known
to the attacker.

\vspace{0.04in}
\textit{\textbf{An attacker's sphere of influence is bounded.}} For
physical sensor threats an adversary is usually assumed to have access to the
physical medium used by the sensor. However, they may not always be within
proximity; they may be limited by their equipment or other environmental
factors. Similarly, an attacker may be temporally limited.


All these assumptions correspond to stronger adversaries and realities of CPS attacks.

\section{\pname \label{sec:model}}

\pname{} is a defensive security technique tailored for CPSs. It combines two
orthogonal, but complementary techniques: \r{} and diversification. This
combination in conjunction with the unique properties of CPSs, play off each
other to provide stronger security than either technique on its own. We discuss
the intuition behind \pname{} below.

\textbf{Why \R?} \pname{} takes advantage of a simple and universally applicable
panacea for software problems, \r{}ting. Even among expert users, a \r{} is the
preferred solution for nearly any problem in the computing world. The simple
intuition behind the effectiveness of this approach is that software is tested
most often in its pristine, fresh state as discussed by Oppenheimer et al.
\cite{oppenheimer2003internet} and Ding \cite{dingrecovery}. With respect to the
overall health of the system, the conditions of a \r{} provide a predictable and
well defined behavior.

From the viewpoint of thwarting an attacker, the restoration of state, whether
it be code, data, or configuration typically helps prevent an attacker's ability
to corrupt the system. For example, simple \r{}s can remove the effects of non
persistent attacks that live in memory. More sophisticated \r{} mechanisms that
may restore code or other information, could protect against persistent threats.
By frequently performing \r{}s, \pname{} limits the effects an attack might have,
as well as, the time an attack has to complete. In other words, an attack has a
bounded time horizon over which it can affect the system, simply because its
effects are frequently removed.

\textbf{Why Diversify?} Typically, once a vulnerability is identified, an
adversary can continuously carry out an attack as long as the vulnerability
remains present. To remedy this, some variability must be introduced into the
system. Otherwise, the same vulnerability would persist. System diversification
introduces randomness to prevent the system from being compromised by the same
method continuously. The benefits of such an approach have been shown to be
successful in a number of related works. As a consequence of diversification,
\pname{} is able to lower the adversary's chance of success.

\textbf{Why \R{} \& Diversify?}
In combination, these techniques provide two main advantages:

The first advantage is that diversification can help protect data that needs to
be carried across \r{}s. Some CPSs may require certain data that cannot be
re-learned during normal operation. For example, sensor calibration data of a
quadcopter can only be obtained while it is not in flight. Therefore, it must
preserved across resets. One mechanism to protect such data is to take advantage
of diversification. Diversification can be used to change the location of the
data on every \r{}, making it harder for an adversary to locate this persisted
data. Another mechanism is encryption. Alternatively, the diversification
strategy may encrypt the data with a rotating key that is changed on every \r{},
making it more difficult to corrupt the data.

The second advantage is that \r{}s can simplify the implementation of certain
diversification strategies. For example, re-diversification on traditional
systems is typically managed in the following two ways. One way involves running
a shadow copy of the program. The shadow copy is diversified in memory and then
swapped in with the original program. This technique works well for applications
such as RESTful---stateless--APIs, because state information does not need to be
shared between the copies. Another diversification mechanism involves a
technique known as taint-tracking that allows pointers and other data structures
to be tracked during execution. Taint-tracking allows for application state to
be migrated between diversified copies of the program. Both of these techniques
add a significant level of complexity in order to achieve re-diversification.
This additional complexity may make it impractical or impossible to do on
certain types of CPSs where resources (e.g RAM) may be limited. With \pname{},
the combination of resets an diversification can help obtain similar results
with less complexity. Since \r{}s bring the program to a known point and the
majority of the application state is discarded, the amount of pointers or other
similar data structures that need to be migrated is significantly reduced.
\pname{} can simply re-diversify and restart execution.

\textbf{Why does this work for CPS?} As discussed in
Section~\ref{sec:system-model}, CPSs have unique properties. We can rely on
\textit{inertia} to survive \r{}s---the system continues operating during a \r{}
even while missing events. We can also rely on the system's \textit{observability}
to re-learn about state of the system, and on \textit{tolerance} to recover from
a \r{} without any intervention. These properties allow us to exploit certain
benefits that make tasks such as diversification and \r{}s simpler and
practical for resource limited CPS.

These scenarios can prove difficult to adapt to traditional IT systems as they
do not necessarily share these properties. Traditional systems would potentially
rely on replication to emulate some of the benefits of inertia. This replication
is usually done at the cost of additional memory or hardware. To account for the
lack in observability, traditional systems might require a secure data store or
some external entity from which to recover it's state.

\textbf{What parameters can we tune?}  We consider \pname{} to have three
distinct modes of operation based on when it chooses to employ \r{}:

\vspace{0.04in}
\textit{Periodic Mode} -- The interval between \r{}s is fixed. For
example, the quadcopter \r{}s every 1-2s.

\vspace{0.04in}
\textit{Random Mode} -- The interval between \r{}s is randomly picked
from a predetermined range that is considered to be safe.

\vspace{0.04in}
\textit{Adaptive Mode} -- The interval between \r{}s is dependent on a
certain set of criteria. In this mode the system or an observer monitors the
effects of a \r{} and selectively chooses when to execute the \r{}. These
criteria are covered in the safety analysis in
Section~\ref{subsec:model-safety}.

These modes of operation have their own security strengths and
weaknesses. Another parameter to tune is the diversification strategy.
Although \pname{} does not specify a particular diversification
strategy, proper selection impacts the strength of \pname. Analysis
on these topics are provided in Section~\ref{sec:model-analysis}.

\section{\pname{} Analysis \label{sec:model-analysis}}

In this section we take a closer look at analyzing \pname{}
specifically with respect to its security and safety.
For the analysis of \pname{}, we first consider the periodic mode and then
expand the arguments to other modes.

\subsection{Security \label{subsec:model-security}}

\begin{figure}[t] \centering
  \includegraphics[width=0.7\linewidth]{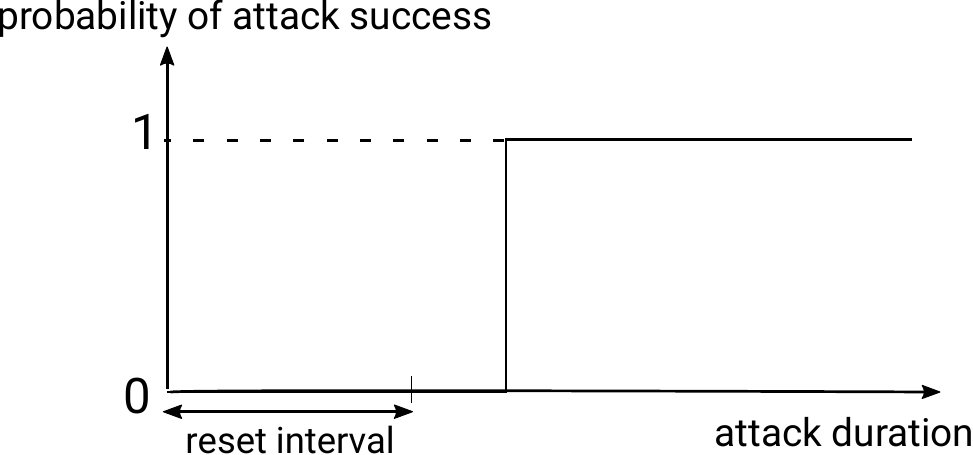}
  \caption{Success probability of a deterministic attack. The probability
remains zero until the adversary collects enough information but goes up to one
after that.}
  \label{fig:prob-b}
\end{figure}

\begin{figure}[t] \centering
  \includegraphics[width=0.7\linewidth]{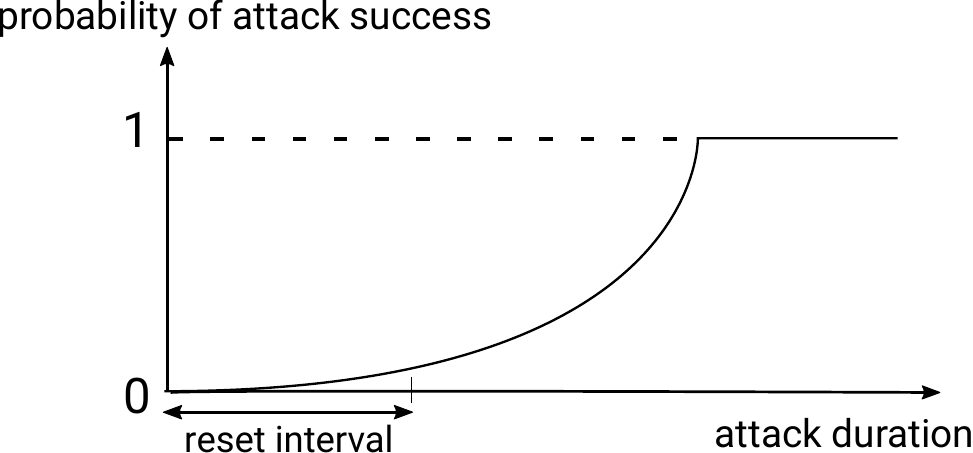}
  \caption{Success probability of an probabilistic attack. The probability
gradually goes up as the adversary tries different guess for a successful
attack. It will eventually reach one if the search is done exhaustively.}
  \label{fig:prob-a}
\end{figure}

\textbf{How does the platform affect security?} For microcontroller-based CPSs,
the platform alone can impose certain limitations on the attacker's capabilities
especially when considering the time bound imposed by frequent \r{}s. To achieve
persistence on these devices that normally execute from Flash, an attacker has
two options: (1) copy the contents of a flash sector(s) to RAM, modify the
contents in memory, erase the sector(s), and finally write the contents of
memory back to flash. (2) Toggle '1' bits to '0' bits in place on flash to
modify the contents.

The steps taken for option 1 take a significant amount of time. To
give some context, the microcontroller used in our case studies
have flash whose smallest sector size is 16KB. Erasing this 16KB
sector takes approximately 210ms and writing (the ``program"
operation) takes around 460ms.  If \pname{} is able to \r{} quickly
enough, an attacker's attempt to achieve persistence can be thwarted.
An attacker may use option 2, i.e., clear all bits to 0 for individual
sectors, to reduce these delays to just the program phase. However,
this option is significantly limited. To understand why, let us say
the attacker wants to change code or data. The only option
they have is to change all of the bits to 0. For instance, if an
instruction is set a jump to \texttt{0x40000}. By writing to the
sector an attacker can change the jump to \texttt{0x00000}. This is
usually an illegal address that causes a memory fault. Further, all
other instructions would be changed to 0x00 in the adjacent bytes.
The same applies to data pages (which need to be programmed for
64KB or larger). So while option 2 may be useful to construct certain
portions of an exploit, it is highly restrictive and unlikely to
be general-purpose. In fact, the only known microcontroller rootkit~\cite{goodspeed2009}
does not use the second option.

\textbf{{How are different attack categories affected?}} We discuss timeliness
more concretely by analyzing the probability of an adversary's success and where
\pname{} fits in. Attacks generally fall within two separate categories
represented by their respective probability functions. The first category, shown
in Figure~\ref{fig:prob-b}, describes an attack that until some condition is met
has zero chance of success. This probability function models a memory disclosure
attack in which a certain portion of memory must be harvested before the
attacker has a chance to hijack the system. Published works describing these
types of attacks report that they require a substantial amount of time. If the
time to harvest memory is longer than the \r{} interval, then \pname{}
successfully defends against the adversary. The second category, shown in
Figure~\ref{fig:prob-a}, describes an attack that over time, slowly becomes more
and more likely to succeed as they exhaust their search space. This probability
function models a guessing attempt to find the location of core data structures
on a system without \pname{}, where given enough time an attacker can eventually
hijack the system. Because \pname{} bounds the duration of an attack, it
essentially restricts the probability of hijack. As long as the \r{} interval
$T_R$ is short enough, then the probability of attack success should remain well
below one. For both of these attack categories, diversification seeks to ensure
that the attacker's previous attempts are invalidated and therefore the
probability of attack success remains close to 0.

\textbf{What known attacks can \pname{} protect against?}
We will discuss two concrete attacks and how \pname{} provides protection. The
first are rootkits. The body of published rootkits for microcontrollers is
limited. However, Travis Goodspeed has done significant work exploring the area
developing rootkits for the MSP430 microcontroller \cite{goodspeed2009}. The rootkit achieves
persistence by writing its payload to flash. Had the system with the MSP430
been reset frequently enough, it might have prevented the rootkit.

A second example of a threat which \pname{} can provide relief are
defeat devices. These defeat devices have been recently brought
into the spotlight after the Volkswagen incident. These devices
gather data over time for a variety of parameters such as engine
runtime and wheel rotation. They then determine how they should
modify their behavior according to whether they are being inspected
or not. Because \pname{}'s resets essentially clear the data that
these ECUs are capturing, it can prevent them from determining the
conditions of an emissions test.  Of course if the adversary knew
about the resets they may engineer more cleverer defeat devices.

\textbf{What are appropriate diversification techniques?} \pname{}
does not specify the diversification technique that must be used.
Because CPS applications have tight timing requirements, a
diversification technique appropriate for one system may not be
applicable for the other. Furthermore, resource constraints of a
particular system may limit that techniques can be implemented.  An
example of an applicable diversification strategy candidate relies
on execution path randomization similar to Isomeron
\cite{davi2015isomeron}. Such a strategy can be implemented at
compile time with limited runtime support making it appropriate for
embedded devices. A compile time Isomeron variant can additionally
support \r{}s. For example, a snapshot of RAM state can be restored
from a known point without the need of patching addresses, or other
variables as the execution of the program still remains diversified.

\subsection{Safety \label{subsec:model-safety}}

The goal of this section is to develop a framework for when \pname{}
can be applied to CPSs safely. To do so, we (1) describe how safety
requirements are determined, (2) determine what conditions are
necessary to fulfill the requirements, and (3) define a set of
parameters to relate the effects of \pname{} on safety.

\begin{figure}[t]
\centering
\def\svgwidth{\linewidth}
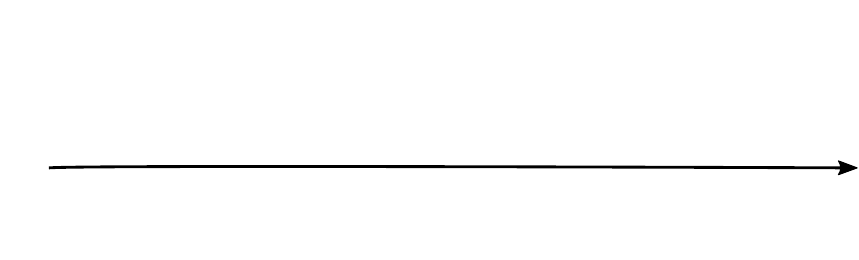
\small
 \begin{tabularx}{\linewidth}{cX}
\\
$T_R$ & The \r{} interval. This is fixed for periodic mode.\\
 $d_R$ & The duration of the \r{}. During this time, there will be no output.\\
 $d_S$ & The duration from the first output to where the system enters a stable state. \\
 $d_{SS}$ & The duration of the stable state  that must be maintained for safety until the next \r{}.\\
 \end{tabularx}

\caption{Intervals between \r{}s. Each interval between \r{}s consists of the
time the system spends in rebooting, stabilizing, and stablized state.}
  \label{fig:safety}
\end{figure}

From Figure~\ref{fig:safety}, the time which the system needs to
recover from one \r{} should be a sum of the downtime of the system
where no output is generated ($d_R$) and the time the system needs
to stabilize ($d_S$). This is due to the typical nature of algorithms
used in CPSs that may require some observation of the environment
to reach a stable state. From Figure~\ref{fig:safety}, we can clearly
see that the following equation holds: \begin{align*} T_R = (d_R +
d_S + d_{SS}) \end{align*} Therefore, the condition to consider is
simply: \textit{Two consecutive \r{}s must be separated by more
than or equal to $d_R + d_S + d_{SS}$.} In other words, $T_R$ is
the parameter that affects the safety of the system.

The above condition can also be interpreted as the ratio of the duration of time
spent in the stable state to the time required to recover to a stable state
after a \r{}. The ratio $D$ can be expressed with the previously defined
parameters as follows:
\begin{align*} D = \frac{d_{SS}}{(d_R + d_S)}
\end{align*} Time spent in the stable state pushes the system into a safe
region, where as, time spent in recovery is time in which the system is not
producing outputs. As a result, the higher the ratio $D$, the safer and more
performant the system can be considered. Given that the downtime $d_R$ and the
time until the first output $d_S$ are hard to control, the system designer's
only practical option is to tweak the duration of being in the stable state
$d_{SS}$. The ratio $D$ is used in Section~\ref{sec:evaluation} for our
evaluation of safety.

\textbf{How do the operation modes satisfy safety?} The three modes of
operation each affect safety differently. The Periodic and Random modes, provide
similar performance to the original system given that $T_R$ is chosen
appropriately. As $T_R$ decreases past some threshold determined by the minimum
time in the stable state, $d_{SS}$, the system stops functioning correctly. The
key for these modes is to provide the minimum $T_R$ while still providing
comparable safety to the system without \pname. The benefit for these approaches
is their simplicity, especially in terms of implementation.

The adaptive strategy can provide the closest performance to the normal system
by continuously monitoring its effects and is specifically meant to address
safety level S2. If \pname{} considers that its actions
will violate any of the safety requirements, $T_R$ can be continuously vary
between an upper and lower bound. The difficulty with this strategy is
determining the appropriate metric to monitor which varies from system to
system.

\textbf{How do we include diversification?} While \pname{} does not
define a particular diversification strategy, one feature they all
have in common is increased work. This can result in additional
delays compared to the baseline system.  In order to satisfy the
safety requirements, the additional overhead introduced by
diversification should still satisfy the original deadlines. Formally,
determining the satisfiability of these timings is done by scheduling
analysis.  Difficulties in accurately modeling the system make this
approach complex leading to experimentation in practice.  Due to
the highly specific nature of safety, further in depth discussion
on the subject is presented when evaluating our case studies in
Section~\ref{sec:evaluation}. For this evaluation, we will demonstrate
appropriate safety requirements and explore the limits of the \r{}
interval $T_R$.
\section{\pname{} Implementation \label{sec:case-studies}}

In order to evaluate \pname{} we study two distinct CPSs: an Engine Control Unit
(ECU) and a UAV flight controller (FC). Each case study provides its own
challenges to determining the feasibility of \pname, because each case is
different with respect to the physical component under control.

\subsection{Engine Control Unit}

An ECU is the brain of an engine, designed to directly process inputs from a
series of sensors and supply output signals to actuators to control the process
of internal combustion. As is common to CPSs, an ECU must perform a set of
real-time tasks for the engine to perform properly. For a combustion engine to produce the
right amount of power, it must inject fuel into its internal chamber, mix it
with air, and finally ignite the air-fuel mixture, all at the right timings.
Typical engines perform these steps in what is called the four-stroke cycle.

\textbf{How it works:} There are two rotating parts, the crank and camshaft,
inside an engine. The ECU observes their revolutions to determine what the state
of the engine is. The number of input signals that must be observed to correctly
determine the engine state depends on the shapes of the crank and camshaft.
Then, the control algorithm interpolates the time to properly schedule ignition
and injection events.
Once the ECU has determined in which phase of the combustion cycle the engine is
in, it will use other measurements from sensors, such as throttle position,
temperature, pressure, air-flow and oxygen to accurately determine the air-fuel
mixture to be injected.

\textbf{Platform:} For our case study we use the rusEFI open-source ECU and a
Honda CBR600RR engine, a very commonly engine used by FSAE racing enthusiasts.
The source-code is written in C/C++ running on top of an open-source real-time
library operating system called ChibiOS and is designed to run on a
STM32F4-Discovery board, a widely popular micro-control unit (MCU). This board
contains a 168 MHz ARM Cortex-M4 processor with 192 KBytes of SRAM and 1MB of
non-volatile flash memory. The MCU contains only a Memory Protection Unit (MPU)
with 8 protection regions, which may be leveraged by diversification techniques.

\textbf{Reset Strategy:} Realizing \pname{} involves selecting an appropriate
\r{} strategy. For the ECU, we choose to power cycle the MCU, which effectively
clears out all hardware state. Simple power cycling, or reboots, provide strong
security advantages as it can be triggered externally, without any software.
Additionally, it protects against attacks which may freeze the configuration of
certain hardware peripherals. Power cycling incurs certain costs, specifically
the cost of rebooting the chip and the time for the startup routines to
reinitialize the controller. However, we found that the cost of rebooting the
chip was on the order of microseconds and thus completely inconsequential
compared to the latency of the startup routine. The non-interactive version of
rusEFI's startup time was 20ms ($d_R = 20\text{ms}$.) This is still very fast
compared to Desktop systems.

\textbf{Diversification Strategy:}
We implement a static variant of Isomeron that provides execution path
randomization. The original implementation of Isomeron uses dynamic binary
instrumentation techniques. This approach is not feasible on resource
constrained devices. By leveraging existing \texttt{BinUtils} functionality
compiler flags, our implementation makes it suitable for our
targeted devices.

\subsection{Flight Controller}

Similar to the ECU, the flight controller is the brain of an aircraft and is
designed to ensure its stability and control. An aircraft has six degrees of
freedom: translation along the \textit{x, y, z} directions and rotation about
the \textit{x, y, z} axes. Each rotational axis is commonly referred to as
\textit{pitch, roll, yaw}, respectively, while the three together are refereed
to as the attitude. Proper attitude control of the aircraft is critical for its
stability.

\textbf{How it works:} The flight controller is primarily responsible for
ensuring attitude stability while aiding a pilot or performing autonomous
flight. It must read all of the sensor data and filter the noise in order to
calculate proper output commands to send to its actuators. In particular, we
focus on quadrotor helicopters more commonly referred to as quadcopters.
Controlling these quadcopters involves operating four independent rotors to the
provide six degrees of freedom. Sensors measuring a number of physical
properties are then fused together to estimate the position and attitude of the
quadcopter. This estimation, similar to the case of ECU, require a certain
number of observation samples before an output is produced. This output is then
used by other components that determine the best command actions for the system.

\textbf{Platform:} For our case study we use the PX4 open-source flight
controller with a DJI F450 quadcopter air-frame, a very common DIY kit favored
by enthusiasts. The PX4 flight controller provides attitude and position control
using a series of sensors such as GPS, optical flow, accelerometer, gyroscope,
and barometer. The PX4 controller software includes a variety of flight modes
ranging from manual, assisted, to fully autonomous. The source-code is written
in C/C++ and supports multiple kinds of OS and hardware targets. Specifically,
we use the Pixhawk board based on the same series of MCU as that used in the ECU
case study. The overall PX4 architecture uses two main estimators corresponding
to the six degrees of freedom: position estimator and attitude estimator. The
estimated values are passed to the position controller and attitude controller
which are then used to compute the optimal trajectory and trust vectors for the
quadcopter. The thrust vectors are then converted from their normalized state to
their raw (PWM---pulse-width modulation) values by a mixer and the result is
directly supplied to the actuators. Depending on the flight mode, certain
components function differently. For assisted mode, pilot inputs are fed
directly to the attitude controller to control the quadcopter, while for
autonomous mode the system is controlled by a navigator which feeds coordinates
to the position control.

\textbf{Reset Strategy:} Similar to the ECU case study, we first attempted
simple reboots. The reboot time $d_R$ for PX4 was found to be around 1.5s. Given
the more sensitive physical dynamics of the quadcopter, simple rebooting is not
effective, i.e., the quadcopter crashed very often, prompting the need of a more
efficient approach. As a result, we implement an optimized \r{} strategy. We found
that much of the startup time was spent in initializing data structures and
setting up the system for operation. So, we create a snapshot right after all
the initialization and use it to practically instantly start the system. This
provides certain security benefits as the snapshot can be verified and signed,
limiting, or even completely eliminating the possibility of tampering from an attack.

This \r{} strategy takes a snapshot of the entirety of RAM. It is then stored in
a special region of flash and at the following boot, the saved state is
restored. The special flash region is protected, and locked by the MPU. This
provides a consistent restoration point for the system's lifetime. This \r{}
strategy was implemented as an extension of the NuttX library operating system used
by the Pixhawk PX4 target. This approach takes approximately 3ms (i.e $d_R =
3\text{ms}$) to restore and is primarily dominated by the time required to write
data from flash to RAM.

When the snapshot is taken, and what data is stored in the snapshot, have
implications on the capabilities of the system. Depending on the flight mode for
the quadcopter the snapshot has different requirements as to what data can be
reset and persisted. For the autonomous flight mode for example, coordinates
for the quadcopter's flight path can be part of the snapshot. However, this
would prevent the quadcopter's flight path from being modified mid-flight. If
this capability is desired, the said data would need to be persisted across
\r{}s. The assisted flight mode has fewer limitations. For the assisted flight
mode, which only requires the pilot inputs, a simple snapshot of the system
taken after the sensors have been calibrated is sufficient, as the system can
recover the state that it needs by reobserving the environment through its sensors.

We found that regardless of the flight mode being used, the snapshot could be taken
once in a controlled and secure environment, as long as, the system was
initialized with the correct parameters. For the assisted mode this meant, the
sensors need to be calibrated, and for the more advanced autonomous mode,
mission waypoints for navigation need to be initialized with absolute values as
relative measurements become a problem unless they are designed to be persisted.
Neither method degraded the quality of the flight.

\textbf{Diversification Strategy:}
For the flight controller we chose to implement an alternate diversification
strategy to show the flexibility of \pname{}. We use a simpler re-randomizable stack
canary strategy. On each reset, we basically randomly generate a new canary.
\section{Evaluation \label{sec:evaluation}}

The most critical component of \pname{} is resets. We require careful evaluation
of its effects on systems.
The main questions we study are: what is the frequency of reset at
which the system becomes unsafe (a) for the ECU and (b) for the flight
controller. Similarly, we also study how the stability of system is impacted for
different frequency modes. In terms of the parameters setup in an earlier
Section~\ref{sec:model-analysis}, we will quantify $d_R$, the reset downtime, and what factors affect
it, as well as, determine realistic values for the \r{} interval $T_R$.
Fundamentally, $d_R$ value is determined by the \r{} strategy used while $T_R$
is dependent on the system physics.

\textbf{Engine Control Unit:} The \r{} time of our ECU is 20ms ($d_R =
20\text{ms}$). The stabilization time $d_S$ for the ECU is dependent on the
number of engine cycles that must be observed. The ECU must observe two engine
cycles to determine whether it is synchronized with the engine's rotation.
Additionally, it must observe enough engine cycles to compute properties that
must be integrated over time (eg. acceleration requires three engine cycles).
Assuming an engine speed of 4500RPM (i.e., approx 75Hz), each engine
cycle takes 13ms, therefore $d_S \approx 39\text{ms}$.

\textbf{\textit{Diversification:}}
For the ECU we studied the effects of our Isomeron implementation. The results
showed our version introduced a constant slow down of approximately 2.13x,
primarily due to its use of a hardware random number generator. While this slow
down may seem large, the original application had more than sufficient slack to
accomodate. To put into perspective, even with the slow down, our ECU was still
within typical timing accuracy of commercial systems. We found that
diversification on the engine had no observable effect.

\begin{figure} \centering
    \includegraphics[width=\linewidth]{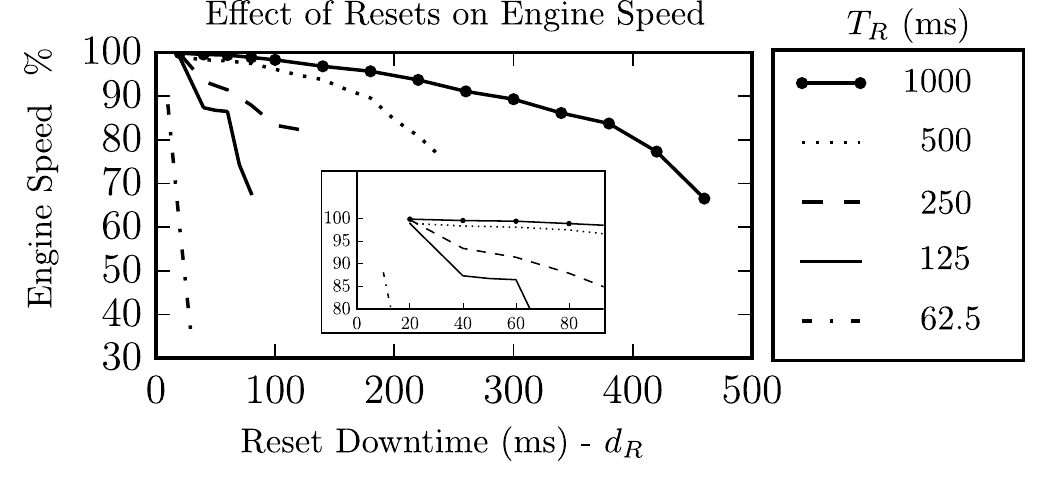}
    \caption{A sweep of the \r{} interval $T_R$ and \r{} downtime $d_R$ to study
the effects on engine speed. We observe that for certain combinations of $T_R$
and $d_R$ the engine speed approximates 100\%.}
    \label{fig:engine-sweep}
\end{figure}

\textbf{\textit{Safety Requirements:}} In order to validate the safety of the
system, we must define feasible safety requirements. We define two such
requirements: (1) \textit{The engine should maintain its speed (i.e. 4500 RPM or 75
Hz).} (2) \textit{The engine should not stop.}

We performed a set of experiments on a real engine to explore the cost of \r{}
as measured by the drop in engine speed, and how it varies with different \r{}
periods ($T_R$) and reset downtimes ($d_R$).
To determine the satisfiability of the first requirement, a sweep of $d_R$ and
$T_R$ are performed given a nominal engine speed of 4500RPM.
Figure~\ref{fig:engine-sweep} shows the change in engine speed as a percentage
for the sweep. Each line in the graph represents a different \r{} interval
$T_R$. We plot 1s, 500ms, 250ms, 125ms, and 62.5ms reboots. From Figure~\ref{fig:engine-sweep}
we can see that the first requirement, maintaining the engine speed, can be
satisfied for a wide range of $d_R$ and $f_R$ where the engine speed is
approximately 100\%.

The second requirement, keeping the engine from stopping, involves the ratio
($D$) of the time the engine spends in its stable state (igniting and injecting
fuel) and the time it spends \r{}ting. We observe that as the ratio $D$
decreases for a fixed $T_R$, the engine speed decreases. At some point,
depending on the ratio $D$, the ECU is not able to generate enough energy to
overcome friction, and the engine comes to a stop, in which case the safety is
violated. We refer to the specific engine speed at which this failure occurs as
the stopping threshold or the minimum $d_R$. As we reboot more frequently, we
observe lower engine speeds without crossing the stopping threshold during
operation. We also note that the actual stall threshold varies non-linearly with
$T_R$ and $d_R$, most likely due to environmental factors and the large
variability in the internal combustion process.

For our experiments, we can therefore conclude that there are specific
combinations of reset periods and reset downtimes for which safety can be
satisfied even as the system misses events. We further realize that perhaps the
most important factor is the ratio D as a result of the difference in time
scales between resets and physical actuation.

\textbf{Flight Controller:} The \r{} strategy implemented for the Flight
Controller takes approximately 3ms to restore the snapshot (i.e. $d_R$ is
$3\text{ms}$). We will perform the rest of the evaluation with this approach in
mind.

\textbf{\textit{Diversification:}} Similar to the ECU, we sought to observe
any difference in the system's behavior introduced by
diversification. Our simple strategy in the case of the quadcopter, had a
negligible effect.

\textbf{\textit{Safety Requirements:}} Determining appropriate safety
requirements for the quadcopter is different from the ECU as different flight
modes may call for different requirements. Similar to the ECU, we define two
safety requirements for our evaluation: (1) \textit{The quadcopter should not
oscillate during flight. In other words, it's attitude should be stable.} (2)
\textit{The quadcopter must not crash and fall out of the sky.}

These two requirements are critical to the safety of the quadcopter as
oscillations limit the control and stability of the system, especially when
attempting to hover. Additionally, if the quadcopter falls out of the sky, then it could
cause irreparable damage to itself and others.

\begin{figure} \centering
    \includegraphics[width=0.7\linewidth]{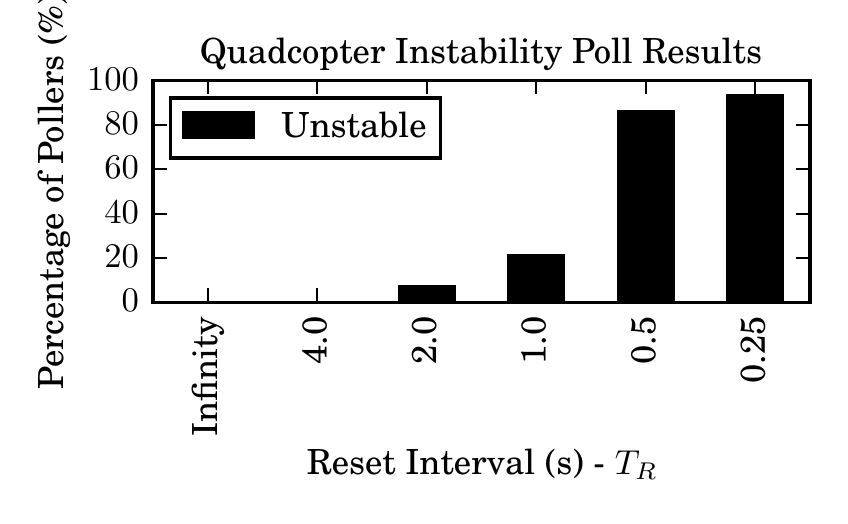}
    \caption{The results of the poll conducted to determine at which \r{}
interval ($T_R$) the quadcopter will start to become unstable during hover. From
this we can determine a minimum $T_R$ for the system. \label{fig:quad-poll}}
\end{figure}

To better gauge the threshold at which a pilot would begin to detect these
oscillations, or in other words, the lower limit for $T_R$, we conducted a survey
among a set of 20 students. The survey was conducted using an ABX test
methodology where various videos of the quadcopter with \pname{} during flight
for different $T_R$ were shown. Before conducting the survey, users were shown
an example video of a stable and unstable flight. They were then shown
videos in a random sequence and asked to
determine whether there were any observable oscillations during hover flight.
The results are shown in Figure~\ref{fig:quad-poll} indicating that oscillations
become significantly observable somewhere between $T_R$ of one-half and one second.

\begin{figure} \centering
  \includegraphics[width=0.7\linewidth]{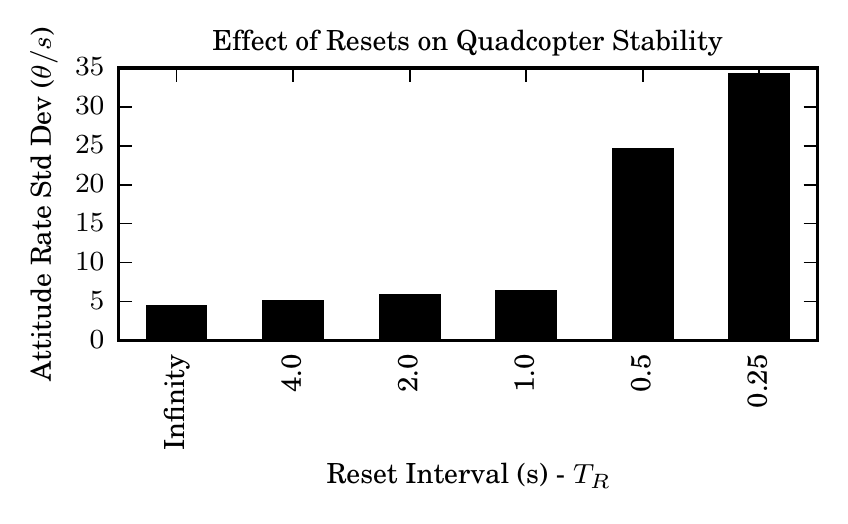}
  \caption{The effects of \r{}s on the Quadcopter Stability for various \r{}
intervals. These results, quantify our observations from the poll
conducted. \label{fig:quad-r-sweep}}
\end{figure}

Next, we relate the results of the poll to technical parameters of flight,
specifically the attitude. We mounted two Pixhawk flight controllers on the
quadcopter: one for control and the other data acquisition to address the
limitations as discussed in Section~\ref{sec:limitations}. To quantify the
effects of \R{}, the standard deviation of the quadcopter's attitude rate over
time from the flight data used in the polls was used. The results are shown in
Figure~\ref{fig:quad-r-sweep}. The results show little impact on the attitude
for $T_R > 1\text{s}$ and a large spike for smaller values. This indicates that
for $T_R > 1\text{s}$ the stability of the system is roughly equivalent to the
system without \pname{} and thus safety is maintained. At lower $T_R$ periods, we
see a large spike in the standard deviations, which correspond to when we
observe the system to start oscillating.

\begin{figure*}[!t] \centering
  \begin{subfigure}[t]{0.47\textwidth} \centering
    \includegraphics[height=1.3in]{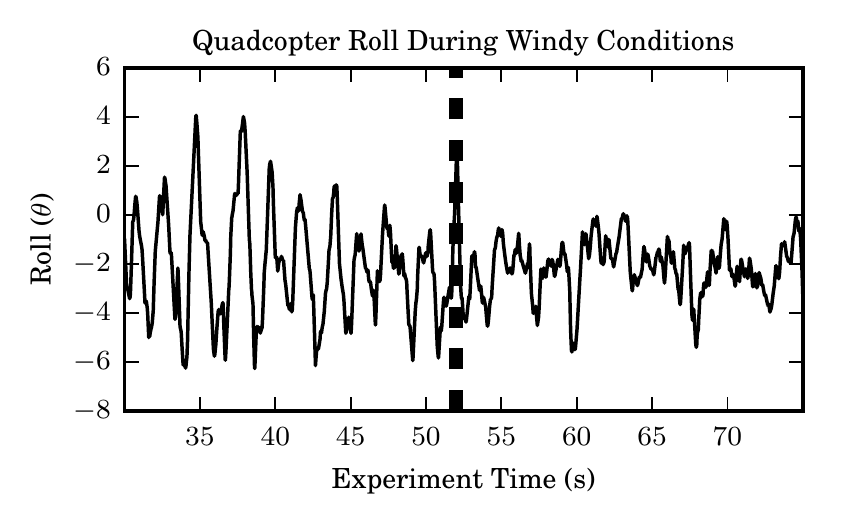}
    \caption{The roll of the quadcopter from \r{} period $T_R$ of 1s (left) to
8s (right) during windy conditions. The dashed vertical line marks the point at
which the adaptive mode switches the period between \r{}s
($T_R$). \label{fig:quad-roll-adap}}
  \end{subfigure} \hspace{1mm}
  \begin{subfigure}[t]{0.47\textwidth} \centering
    \includegraphics[height=1.3in]{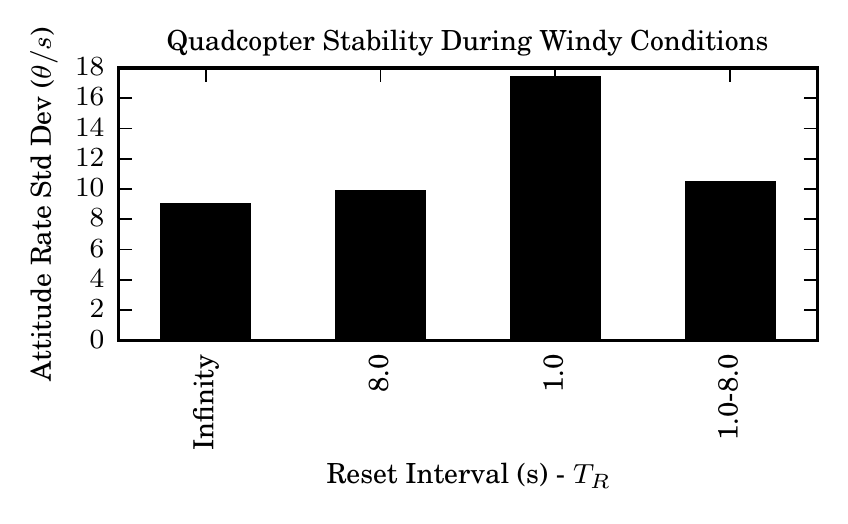}
    \caption{The stability of the quadcopter's attitude during windy conditions.
Shows the comparison between baseline and different $T_R$.}
    \label{fig:quad-att-wind}
  \end{subfigure}
  \caption{Adaptive Operation Mode with wind.}
  \label{fig:ecu-eval}
\end{figure*}

\textbf{\textit{Adaptive Mode:}} Given the variety of external forces a
quadcopter may be subject to, it makes sense for this system to consider the
Adaptive mode to satisfy safety requirements. This would allow the system to
respond to wind, among other external factors. The upper and lower bounds chosen
for this mode, will thus determine the minimum and maximum $T_R$. At worst if we
assume the wind is constant the system will behave no worse than the lower bound
$T_R$. In other words, this case will be equivalent to the Periodic mode at the
given $T_R$. In reality since wind is typically varied, due to gusts, the
average effective $T_R$ of system should fall somewhere between the upper and
lower bounds. To demonstrate this we simulated wind using multiple fans blowing
into the path of our quadcopter. We operate the quadcopter at $T_R = 1\text{s}$
for half of the time and at $T_R = 8\text{s}$ for the other. Given the direction
in which our fans were blowing, Figure~\ref{fig:quad-roll-adap} details the
quadcopter's roll angle as affected by the wind. The dotted line marks the point
where the transition from a faster to slower \r{} period is made. On the left side we
see $T_R = 1\text{s}$ with high fluctuations while on the right we see $T_R =
8\text{s}$. From the results shown in Figure~\ref{fig:quad-att-wind} we observe
that the adaptive mode's performance follows closer to the lower bound $T_R =
8\text{s}$.

\begin{figure}
  \includegraphics[width=\linewidth]{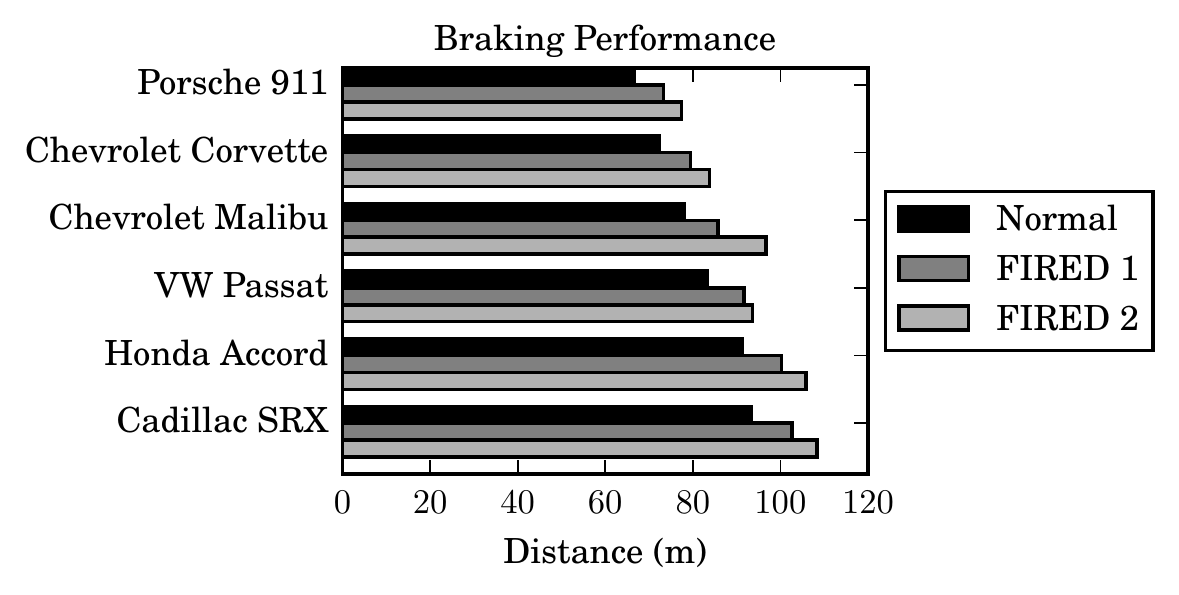}
  \caption{Braking Performance for various cars.}
  \label{fig:braking}
\end{figure}

\textbf{Brake Controller:} We evaluate how \pname{} would affect the braking
distance of a car if deployed on a brake controller. It is
arguable what safety level (S1-S3) would apply for a brake controller. We argue
that it can be categorized as S1 and thus \pname{} applies.
Under ideal dry conditions, a car can usually achieve a deceleration of 8$m/s^2$.
\pname{}'s resets ultimately increase the stopping time of the car, resulting in
a slower deceleration rate. We condiser two pairs of $T_R$ and $d_R$, \pname{} 1
and \pname{} 2 respectively. For \pname{} 1, we assume a reset period ($T_R$) of
one second and a reset duration ($d_R$) of 100ms. The effective deceleration in
this case is roughly 9\% slower or 7.27($m/s^2$). For \pname{} 2, we assume a
$T_R$ of 125ms and $d_R$ of 20ms (the same parameters as the ECU) with an
effective deceleartion of roughly 13.8\% slower or 6.96($m/s^2$).

    Under normal driving conditions the effects of \pname{} are marginal and in
fact are less than the margin of variability seen among different cars in
Figure~\ref{fig:braking}. Additionally, a driver naturally compensates for this
difference in deceleration. It is equivalent to drivng in one car of one make
versus another. There are ways to compensate for these decreases in deceleration
as well. Perhaps the simplest approach is to use tires with better traction.
\section{Limitations, Mitigations and Applications \label{sec:limitations}}

\textbf{Temporary loss of control unacceptable:} In some CPSs even a
temporary loss of control due to resets may be unacceptable. Examples
of these include airplanes during landing or takeoff. Fortunately,
these systems also include multiple replicas for fail safe operation
and interleaved resets will solve the problem. The Boeing 787 is
an example of system that recently recommended interleaved
resets~\cite{boeingAD} for a safety issue. \pname{} is also useful
for autonomous vehicles, such as commodity package delivery drones
where (a) cost is a concern to include redundant components and (b)
temporary loss of control may be acceptable since humans are not
in the drones to experience discomfort.

\textbf{Wear \& Tear:} \pname{} may have miscellaneous effects on CPSs
that may not have been observable from our current evaluation
metrics. These effects can include additional wear and tear of
components for example. It is difficult to say whether \pname{} may
have long term effects as well. However, after having explored these
systems, the physical subsystems are built
with ample tolerance margins such that we may never see any effects
for the duration of the systems lifetime. For example, the rotors
on quadcopters are Brushless DC motors (BLDC).  BLDCs rely on
electronic commutation as opposed to mechanical commutation and
therefore do not suffer from much wear and tear.

\textbf{Log operations:} It is possible that some CPSs include
logging operations that require writing to Flash. In our
model we reset before a write to a flash sector can complete
thereby denying permanance. To enable logging the CPS may
have to be architected to whitelist certain writes to
special write only devices.

\textbf{A drip-drip attack:} While we prevent complete reprogramming
of any flash sector, it is possible that an adversary perform a partial
reprogramming to somehow get the flash memory to hold code or data
they want. This type of attack is theoretically possible, but practical
issues such as controlling the voltage of the flash, or obtaining
enough writes/sectors are likely to hinder the attacker.

\newcolumntype{R}[2]{%
    >{\adjustbox{angle=#1,lap=\width-(#2)}\bgroup}%
    l%
    <{\egroup}%
}
\newcommand*\rot{\multicolumn{1}{R{45}{1em}}}

\section{Related Work}

\begin{table*}[t]
  \footnotesize
  \setlength{\tabcolsep}{5.7pt}
    \caption{Comparison of related works.}
    \label{tab:literature}
    \begin{tabular}{@{}llllllllllllllllllllllllll@{}}
    \toprule
      \footnotesize
    & \multicolumn{13}{c}{General Security Technique}
    &
    & \multicolumn{6}{c}{CPS Security Technique}
    \\
    \cmidrule{2-14} \cmidrule{16-21} 
    \\
     & \rot{Formal Method \cite{klein2009sel4}}
     & \rot{Type Safe Language \cite{hickey2014building}}
     &\rot{Static Analysis \cite{wang2012improving}}
     &\rot{Control Flow Integrity \cite{abadi2005control}}
     &\rot{Execute Only Memory \cite{braden2016leakage}}
     &\rot{Static Layout Randomization \cite{jackson2013diversifying}}
     &\rot{Dynamic Layout Randomization \cite{bhatkar2008data}}
     &\rot{Layout Re-Randomization \cite{bigelow2015timely,williams2016shuffler}}
     &\rot{Control Flow Randomization \cite{davi2015isomeron}}
     &\rot{Intrusion Detection \cite{roesch1999snort}}
     &\rot{Attestation \cite{seshadri2004swatt}}
     &\rot{Malware Removal \cite{kim2010intrusion}}
     &\rot{Proactive Recovery \cite{castro2002practical}}
     &
     &\rot{Sensor Authentication \cite{shoukry2015pycra}}
     &\rot{Spoofing Detection \cite{pasqualetti2013attack}}
     &\rot{Measurement Set Randomization \cite{rahman2014moving}}
     &\rot{Impact Mitigation \cite{urbina2016limiting}}
     &\rot{Resilient State Estimation \cite{fawzi2014secure}}
     & \rot{\pname{} (\textit{this paper})}
     \\
    \midrule
    \textbf{Attack}
    \\
    Cyber
      & \checkmark
      & \checkmark
      & \checkmark
      & \checkmark
      & \checkmark
      & \checkmark
      & \checkmark
      & \checkmark
      & \checkmark
      & \checkmark
      & \checkmark
      & \checkmark
      & \checkmark
      & 
      &
      &
      &
      &
      &
      & \checkmark
      \\
    Physical
      &
      &
      &
      &
      &
      &
      &
      &
      &
      &
      &
      &
      &
      & 
      & \checkmark
      & \checkmark
      & \checkmark
      & \checkmark
      & \checkmark
      &
    \\
    \midrule
    \textbf{Defense}
    \\
    SBC 
      & \checkmark
      & \checkmark
      & \checkmark
      &
      &
      &
      &
      &
      &
      &
      &
      &
      &
      & 
      &
      &
      &
      &
      &
      & \checkmark
      \\
    Detection
      &
      &
      &
      & \checkmark
      & \checkmark
      & \checkmark
      & \checkmark
      & \checkmark
      & \checkmark
      & \checkmark
      & \checkmark
      &
      & \checkmark
      & 
      & \checkmark
      & \checkmark
      &
      &
      &
      & \checkmark
      \\
    Graceful D.
      &
      &
      &
      &
      &
      &
      &
      &
      &
      &
      &
      &
      &
      & 
      & \checkmark
      &
      & \checkmark
      & \checkmark
      & \checkmark
      & \checkmark
    \\
    Impersist.
      &
      &
      &
      &
      &
      &
      &
      & \checkmark
      &
      &
      &
      & \checkmark
      & \checkmark
      & 
      &
      &
      &
      &
      &
      & \checkmark
    \\
    \midrule
    \textbf{Deployment}
    \\
      $\mu C$ Target
      & \checkmark
      & \checkmark
      & \checkmark
      &
      &
      & \checkmark
      &
      &
      &
      &
      & \checkmark
      &
      & \checkmark
      & 
      & \checkmark
      &
      & \checkmark
      & \checkmark
      & \checkmark
      & \checkmark
      \\
    No Add. Logic
      & \checkmark
      & \checkmark
      & \checkmark
      & \checkmark
      & \checkmark
      & \checkmark
      & \checkmark
      & \checkmark
      & \checkmark
      & \checkmark
      &
      & \checkmark
      &
      & 
      & \checkmark
      & \checkmark
      & \checkmark
      & \checkmark
      & \checkmark
      & \checkmark
    \\
    \midrule
  \textbf{CPS property}
    \\
    Inertia
      &
      &
      &
      &
      &
      &
      &
      &
      &
      &
      &
      &
      &
      & 
      &
      &
      &
      &
      &
      & \checkmark
      \\
      Predictability
      &
      &
      &
      &
      &
      &
      &
      &
      &
      &
      &
      &
      &
      & 
      & \checkmark
      &
      & \checkmark
      & \checkmark
      & \checkmark
      & \checkmark
      \\
    \bottomrule
    \end{tabular}
\end{table*}

CPS Security can be broadly categorized along four dimensions: the threats
covered by the defense, the nature of the defense, suitability of the defense to
typical CPS environments, and what specific CPS properties, if any, are used to
build defenses. Table~\ref{tab:literature} presents security techniques relevant
to the paper from general security techniques to those specific to CPSs.The
references used here, especially for the general-purpose security techniques,
are not meant to be exhaustive. These are the most recent and most closely
related to \pname{}.

\textbf{Threat Covered:} Following the duality of the CPS, we categorize the
threats covered into two groups:
(1) \textit{Cyber}---These attacks target the cyber subsystem. This group are those
seen on traditional IT systems.
(2) \textit{Physical}---These attacks target the physical subsystem. More
specifically, it is those attacks which target sensors and actuators.
The general security techniques only focus on cyber attacks whereas CPS specific
techniques primary focus is on physical attacks.

\textbf{Defense Type:} We categorize the defense types into four groups based
when the defensive method is applied (during construction or deployment; and
what properties are provided by the defense under attack):
(1) \textit{Secure-by-Construction (SBC)}---These defenses focus on preventing the root
cause for vulnerabilities.
(2) \textit{Detection}---These defenses focus on detecting system exploitation during
deployment. Does not involve response to a detected attack.
(3) \textit{Graceful Degradation}---These defenses focus on mitigating the malicious
effects of an attack to maintain acceptable levels of performance or to prolong
the life of the system.
(4) \textit{Impersistence}---These defenses focus on denying the attacker an ability
to gain a foothold or getting rid of foothold the attacker may have gained.

Off-line techniques (formal methods\cite{klein2009sel4}, type safe
languages\cite{hickey2014building}, static analysis\cite{wang2012improving}) are
secure-by-construction. These techniques are ideal for low power embedded
systems since they do not pose any restrictions at run-time. However, we have
not seen wide spread use of these techniques in CPS development yet, except for
very critical domains.

On-line security techniques focus on attack detection. While the general
security techniques focus on software intrusion detection, the CPS specific
techniques focus on detection of attacks on physical interfaces, namely sensors
and actuators. \pname{} fits into this category by making attacks that exceed
the \r{} interval impossible.

Graceful degradation under physical attacks has been an active research area in
CPS security. The main idea here is to continue operating, perhaps suboptimally,
even under attack. In most of these papers, the threat that is considered is
an availability attack, and thus continued operation is synonymous with security.
Recently proposed techniques
\cite{rahman2014moving,urbina2016limiting,fawzi2014secure} are approaching a
point where non-detectable levels of destruction can only affect the system so
slowly that stealthy spoofing attacks will become negligible or can be coped
with in practical ways.

Impersistence is one of the crucial aspects in CPS security in that it is almost
impossible to re-install the firmware during their operation. Moreover, it is
essential for many CPS for recovery be performed without human intervention. One
of the closest techniques to \pname{} among general security techniques is
proactive recovery\cite{castro2002practical}. Their idea is similar to ours but
assumes redundancy (i.e. additional hardware) in place of inertia.


Safety of a \r{} in the CPS context has been discussed
before\cite{abdi2016reset}. In fact, for the sake of safety, most of the CPSs
require the ability to survive \r{} during operation, given that a malicious
attack or hardware fault can cause the controller to crash at any time. The
novelty of our work in this context is that by leveraging the CPS properties,
we can survive resets on systems with no additional hardware.

\textbf{\textbf{Deployment:}} CPSs are diverse and so is the computing hardware
used for them. While some are as powerful as servers used in traditional IT
systems, many have only limited processing power. In order to evaluate whether a
technique can be widely applied to CPS, we consider two items:
(1) \textit{Microcontroller Target}---As an extreme case whether the technique can be
deployed, we consider a microcontroller with limited processing
power and memory (e.g. Cortex-M based STM32F4-Discovery
board\cite{stmdiscovery}) that executes in ROM and has no MMU.
(2) \textit{No Additional Logic}---This means no special hardware is necessary for
implementing the technique.

Some of the general techniques cannot be adopted in CPS due to their resource
demand. Most CPSs that go though mass production are resource limited due to
cost. Thus, it is usually unacceptable to increase the computing resources (e.g.
CPU frequency and amount of memory) just for security reasons. On the other
hand, the addition of simple components may be acceptable as its impact on cost
can be negligible. For example, \pname{} can be implemented by using a simple
timer whose configuration cannot be altered from software exposed to attacks.

\textbf{Use of CPS Properties:} The CPS properties that we focus on are:
(1) \textit{Inertia}---This means that a technique relies on the inertia of the
system.
(2) \textit{Predictability}---This means that a technique relies on the fact that
motion of the system can be predicted.

There is no general security techniques that leverage these CPS properties while
most of CPS specific techniques make use of them. To the best of our knowledge,
our work is the first to make use of both inertia and predictability of the
physical subsystem.
\section{Conclusion \label{sec:conclusion}}

It is natural to ask if CPSs are indeed unique, if so, how CPS defenses
should be different from general-purpose defenses. This paper provides one
answer to this question. We construct a new defense that is only practical on
CPSs because of their properties. In contrast to prior work on CPSs, which focus
mostly on sensors, our defense is tailored for the cyber portion of the
CPS. We show that new, simple to implement, low-resource, and effective defenses
are possible if we leverage the unique physical properties of CPS.

We present a new CPS-tailored cyber defense called \pname{} that combines \r{}
and diversification. \pname{} leverages unique properties of cyber-physical
systems such as inertia for its implementation. In a traditional system, frequent
resets will degrade the usability of the system. In CPSs, however, the CPS can
continue to move and operate even during resets because of the momentum/inertia
in the system.

In this paper, we showed that \pname{} is an effective practical defense for an
engine control unit and a flight controller. From our experiments, we determine
that \r{}s can be triggered frequently, as fast as every 125ms for the ECU and
every second for the flight controller, without violating safety requirements.

The security benefits of \pname{} are two-fold: (a) the resets deny persistence
to the attacker. Each reset wipes all volatile or corrupted state, and any
write to persistent, non-volatile storage is denied because resets happen too
frequently to complete a write to non-volatile memory (b) resets can be used to
amplify the security offered by some diversification techniques.

The results of our work show that \r{}s, which may have been previously thought
of as unrealistic due to safety, can indeed be done without violating safety
requirements. When applicable, \pname{} may be especially useful for emerging
unmanned CPSs such as drones.

{\footnotesize \bibliographystyle{acm}
\bibliography{main.bib}}

\theendnotes

\end{document}